\def\lastrev{Revised: WJS 16 Nov 98}

\documentstyle{mn}

\def\ltsima{$\; \buildrel < \over \sim \;$}
\def\simlt{\lower.5ex\hbox{\ltsima}}
\def\gtsima{$\; \buildrel > \over \sim \;$}
\def\simgt{\lower.5ex\hbox{\gtsima}}

\def\apj{ApJ}
\def\mnras{MNRAS} 
\def\apjs{ApJS}
\def\nat{Nat} 

\def\hmpc {\,h^{-1}{\rm Mpc} }
\def\hmpcc {\,h^{-3}{\rm Mpc}^{3} }
\def\Phat {\widehat{P}}
\def\Pe {P_{\rm e}} 
\def\Pshot{P_{\rm shot}} 
\def\nbar {\overline{n}} 
\def\kunit {\, h \,{\rm Mpc}^{-1}}
\def\omegalam {\Omega_\Lambda} 
\def\kms {{ \, \rm km \,s^{-1}}}
\def\ten#1{ \times 10^{#1} }
\def\etal {{\em et al.}} 
\def\bk {{\bf k}} 
\def\br {{\bf r}} 
\def\sumcel{\sum_{\rm cells}} 

\def\plotfiddle#1#2#3#4#5#6#7 { 
 \vskip 0.1truein 
  \vskip #2 
 \includegraphics{#1} }

\def\colhead#1 {#1}   

\title[Power Spectrum of PSCz] 
 {The Power Spectrum of the PSC Redshift Survey} 

\author[W. Sutherland \etal] 
       { W. Sutherland$^{1,2}$, 
	H. Tadros$^{3,1}$,
	G. Efstathiou$^{4}$,
	C.S. Frenk$^{5}$,
	O. Keeble$^{6}$,
	S. Maddox$^{4}$, \newauthor  
        R.G. McMahon$^{4}$,
        S. Oliver$^{6}$,
	M. Rowan-Robinson$^{6}$,
        W. Saunders$^{7}$,
	S.D.M. White$^{8}$ \\
$^1$ Dept. of Physics, Keble Road, Oxford OX1 3RH, UK \\
$^2$ Visiting Observer, Cerro Tololo Interamerican Observatory \\
$^3$ Astronomy Centre, University of Sussex, Brighton BN1 9QH, UK \\
$^4$ Institute of Astronomy. Madingley Road,     Cambridge CB3 0HA, UK\\
$^5$ Dept. of Physics, South Road, Durham DH1 3LE, UK \\
$^6$ Astrophysics Group, Imperial College,  Blackett Laboratory, 
     Prince Consort Road, London SW7 2BZ, UK \\
$^7$ Royal Observatory, Blackford Hill, Edinburgh, EH9 3HJ, UK \\
$^8$ Max Planck Institute for Astrophysik,  Karl-Schwarzschild-Strasse 1,  
   D-8046 Garching bei Munchen, Germany.
	}  

\date{\lastrev} 

\begin{document}

\maketitle  



\begin{abstract} 
\rightskip = 0.0in plus 1em
We measure the redshift-space power spectrum $P(k)$ 
for the recently completed 
IRAS Point Source Catalogue (PSC) redshift survey, 
which contains 14500 galaxies 
over $84\%$ of the sky with 60 micron flux $\ge 0.6$ Jansky.
Comparison with simulations shows that our estimated errors on $P(k)$ are
realistic, and that systematic errors due to the finite
survey volume are small for wavenumbers $ k \simgt 0.03 \kunit$. 
At large scales our power spectrum is intermediate between those
of the earlier QDOT and 1.2 Jansky surveys, 
but with considerably smaller error bars;
it falls slightly more steeply to smaller scales. 
We have fitted families of CDM-like models 
using the Peacock-Dodds formula for 
non-linear evolution; the results are somewhat sensitive to
the assumed small-scale velocity dispersion $\sigma_V$.
Assuming a realistic $\sigma_V \approx 300 \kms$  yields a shape parameter 
$\Gamma \sim 0.25$ and normalisation $b \sigma_8 \sim 0.75$; 
if $\sigma_V$ is as high as $600 \kms$ then $\Gamma = 0.5$ is only
marginally excluded.  
There is little evidence for any `preferred scale' in the power spectrum 
or non-Gaussian behaviour in the distribution of large-scale power. 
\end{abstract}
\vspace{-5mm}

\begin{keywords}
  {surveys -- large-scale structure of Universe -- galaxies: distances and
   redshifts} 
\end{keywords} 

\section{Introduction}
\label{sec-intro}

As is well known (e.g. Peebles 1980), 
the power spectrum of the galaxy distribution on large scales is
of great importance for testing cosmological models, 
since it can be related to the initial conditions by linear perturbation
theory. 
The power spectrum has been estimated from a variety of galaxy 
redshift surveys, notably 
the CfA redshift survey \cite{pvgh}, the QDOT survey 
(Feldman, Kaiser \& Peacock 1993, hereafter FKP), 
the Las Campanas redshift survey \cite{lin96}, and the 
1.2 Jy IRAS survey \cite{12jy-pow}. 
Also, the real-space power spectrum
has been inferred from the APM Galaxy Survey \cite{be93,be94} by 
inversion of both the angular correlation function and 
2-D power spectrum. 

Despite these substantial surveys, 
there are still considerable uncertainties in the
shape of the power spectrum on large scales, since most of these surveys
contain only a small number of independent structures, while the
largest one (Las Campanas) has a slice-like geometry which complicates
the estimation of the power spectrum. 
If the primordial power spectrum is $P(k) \propto k^n$ 
with $n \approx 1$ as suggested by 
inflation, then for consistency with the COBE DMR results 
the present-day power spectrum must show a turnover to this slope 
at $k \simlt 0.02 \kunit$,  
close to the largest scales accessible to current galaxy surveys. 
There is marginal evidence for such a turnover in the APM data
\cite{be94,apm3,ted}. 

Also, it is valuable to measure the power spectrum from surveys
with different selection criteria (e.g. optical \& IRAS selection). 
This is of considerable interest since the observed
power spectrum is measured from the density field of {galaxies},
whereas theory predicts the power spectrum 
of the {mass} distribution.
The process of galaxy formation is poorly understood, so the
observed $P_g(k)$ may differ from $P_m(k)$, possibly in a complex way; 
indeed, since it
appears that IRAS galaxies and optical galaxies have different
small-scale correlation amplitudes, at least one of these cannot
trace the mass. A simple `linear bias' model is often assumed, 
in which $\delta_g = b \delta_m$ for some constant `bias factor' $b$
which may depend on galaxy type;  this model predicts that $P(k)$ for
optical and IRAS galaxies should differ by a multiplicative
factor of $(b_O / b_I)^2$. 
Such a model is reasonable since 
it has been shown by several authors (e.g. Fry \&
Gaztanaga 1993; Cole \etal\ 1998; Mann \etal\ 1998) 
that if the galaxy density is a (possibly complex and stochastic)
function only of the {\em local} 
mass density on scales $\simlt 1 \hmpc$, 
then the effective bias parameter defined by
$b(r) \equiv \sqrt{\xi_g(r)/ \xi_m(r)}$ tends to a constant on large
scales, so such a `local' bias cannot alter the large-scale shape of
the power spectrum.  

Another motivation for measuring the power spectrum from a large
sample of IRAS galaxies is that there appears to be a marginal 
discrepancy between the power spectra from the previous QDOT 
(Feldman, Kaiser \& Peacock 1993, hereafter FKP) 
and 1.2 Jansky \cite{12jy-pow} IRAS surveys, with the amplitude
of $P(k)$ from QDOT being roughly a factor of 2 higher at large scales.
A counts-in-cells comparison of the two surveys does not reveal
any obvious systematic errors \cite{e95}, but it is interesting
to check whether these differences are consistent with
sampling fluctuations in one or both surveys. 

In this paper, we estimate the redshift-space power spectrum
from a new redshift survey \cite{saunders-wfs} 
of some 14,500 galaxies over $84\%$ of the
sky, selected at $60 \mu$m from the IRAS Point Source Catalog (PSC). 
A variety of analyses from this survey will appear shortly;
the topology of the density field has been analysed by Canavezes \etal\
(1998), the correlation function is analysed by Maddox \etal\ (in
preparation), 
the dipole is estimated by Rowan-Robinson \etal\ (in preparation),
a reconstruction
of the peculiar velocity field is given by Branchini \etal\ (1998), 
and the redshift-space distortions 
are estimated using spherical harmonics by Tadros \etal\ (in preparation).

The plan of this paper is as follows:
in \S2 we summarise the construction and properties of the survey; 
in \S3 we present the power spectrum estimates, and we compare
these with results of N-body simulations in \S4. 
In \S5 we compare our results with other surveys and 
some parametrised cosmological models, and also set limits on
non-Gaussian behaviour and periodicities.  


\section{The PSC Redshift Survey}
\label{sec-pscz}
The construction of the PSC redshift survey (hereafter PSCz) 
is described in detail
elsewhere (Saunders \etal\ 1996; Saunders \etal\, in preparation), 
but we summarise the main points here. 
The aim of the survey is to obtain redshifts for all galaxies with 
 $60\mu$m flux $f_{60} > 0.6$ Jy over as much of the sky as feasible. 
The starting point for the survey is the QMW IRAS Galaxy Catalogue
\cite{qigc}, but with modifications to extend the sky coverage and
improve completeness.  
We have relaxed the IRAS colour criteria for galaxy selection, 
and we have added in additional sources in the `2-HCON' sky as follows: 
the IRAS satellite covered most of the sky with 3 hours-confirmed scans
(HCONs) \cite{psc-supp}, while about 20\% of the sky had only 2 HCONs.
Since a source must be detected in 2 separate HCONS for inclusion, 
the PSC catalogue may be less complete in the 2-HCON regions. 
Thus, in the 2-HCON sky we added sources to our target list 
which had a 1-HCON detection 
in the `Point Source Reject' file  and also had a matching entry in the
IRAS Faint Source Catalog. 

These relaxed selection criteria allowed more contamination 
of the target list by non-galaxy sources, but these were excluded
using APM or COSMOS scans of the POSS and UKST sky survey plates.  
If the APM or COSMOS data showed no `obvious' galaxy candidate near the
IRAS source, we visually inspected the
plate and attempted to classify the source, rejecting it if it 
showed an obvious Galactic counterpart e.g. an HII region, 
planetary nebula, dark cloud etc. 
We also exclude very faint galaxies ($B_J > 19.5$) from the redshift
survey since measuring their redshifts is time-consuming, and they are
usually at $z > 0.1$ and hence have little effect on most of the
desired analyses. 

The sky coverage of the survey is the whole sky, excluding 
areas with less than $2$ HCONs in the IRAS data, regions with optical
extinction $A_V > 1.42$ mag as estimated from the IRAS $100\mu$m maps, 
and two small areas near the LMC and SMC. The resulting
coverage is $84\%$ of the whole sky. (An extension to 
93\% sky coverage is in progress, 
using a combination of K-band snapshots and HI redshifts). 

Our 2-D source catalogue contains 17060 IRAS sources in the unmasked sky. 
Of these, 
1593 are rejected as objects in our own Galaxy (e.g. cirrus, bright stars, 
reflection nebulae, planetary nebulae etc),
or as multiple entries from very nearby galaxies `broken up' by
the IRAS point source detection scheme. 
Another 648 sources are rejected either as 
very faint galaxies ($\sim 400$) or as sources without
an optical identification. 
This leaves 14819 galaxies in the `target' list, and
redshifts are now known for 14539 of these (98\%). 

Of these redshifts, $\sim 6500$ are from a combination
of the 1.2 Jy survey \cite{12jy-data} and QDOT \cite{qdot-data}, 
and  $\sim 3000$ are from other publications and private
communications. A further 4115 redshifts were measured by us 
for this survey, using 49 nights at the Isaac Newton Telescope, 
18 nights at the Cerro Tololo 1.5-meter, 
and 6 nights at the Anglo-Australian Telescope,
between 1992 January and 1995 July.  
Details of the observations and data reduction will be given
elsewhere (Saunders \etal, in preparation). 
The error on our redshifts is typically
$150 \kms$; for the literature redshifts it is somewhat smaller. 
The median redshift of the sample is $\approx 8500 \kms$, though 
there is a long `tail' extending to $> 30000 \kms$ due to
the broad luminosity function of IRAS galaxies. 

\section{Power Spectrum Estimation}
\label{sec-pk}

\begin{figure*}
\vskip 5.0truein
\includegraphics{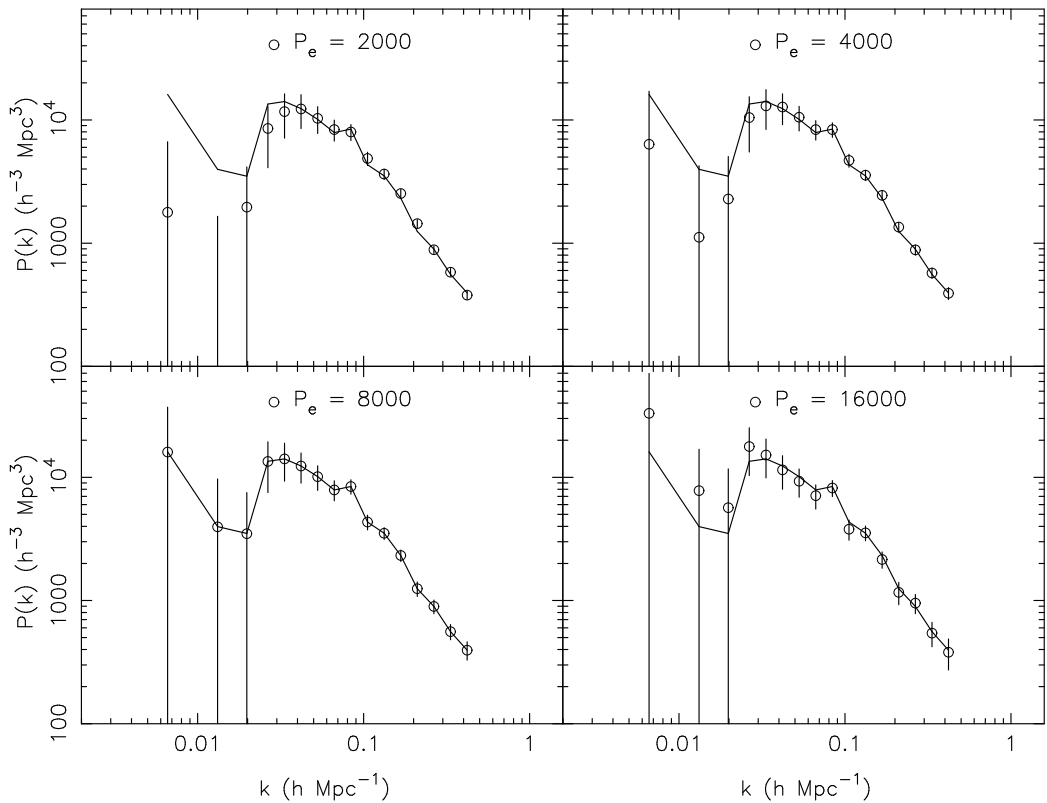} 
\caption{ Estimated redshift-space power spectra $\Phat(k)$ for weights 
 $\Pe = 2000, 4000, 8000, 16000 \hmpcc$. 
 Open circles with error bars show
 the result for each weight, while the solid line 
 is the result for $\Pe = 8000 \hmpcc$ (the same in all panels). 
 \label{fig-pkwt} } 
\end{figure*}

For the estimates here, we restrict the analysis to 
the unmasked sky with the additional constraint 
$\vert b \vert > 10^o$, since the survey may be slightly incomplete
below this latitude; this gives a coverage of $78\%$ of the full sky. 
We also set an upper redshift limit of $cz < 45000 \kms$, 
since the survey is incomplete at high redshift as noted above. 
This gives 13346 galaxies in the `default' sample used for the power spectrum
estimate. 

Since the geometry of the survey is well approximated by a 
sphere, apart from the missing slice near the galactic plane, 
we follow the analysis of Feldman, Kaiser \& Peacock (1994, hereafter FKP) 
with minor modifications.  
This method provides an optimal weighting scheme with redshift
for estimating the power spectrum of an all-sky survey. 
More sophisticated methods have been suggested by e.g. Tegmark (1995); 
these are very useful for surveys with highly non-spherical geometries
but are more complex than necessary for our survey. 

We convert the galaxy positions 
to comoving coordinates assuming $\Omega_0 = 1$ and redshifts
in the Local Group frame, 
and bin the galaxies in a cube of size $950 \hmpc$ with 
$128^3$ cells. 
The FKP method assigns a redshift-dependent
weight to each object, 
\begin{equation}
\label{eq-weight}
  w(r) = {1 \over \sqrt{A} (1 + \Pe \nbar(r))} 
\end{equation}
where $\nbar(r)$ is the mean galaxy density at distance $r$, 
$\Pe$ is the estimated power spectrum (at some scale to be
determined), and $A$ is a normalisation constant (see later). 
We use a parametric fit for the selection function
determined using the method of Springel \& White (1998); 
here this takes the form 
\begin{eqnarray}
\label{eq-selfn} 
 \nbar(z) & = &  n_* y^{(1-\alpha)} / (1 + y^\gamma)^{(\beta/\gamma)}   \\
     y & \equiv & z / z_* , \qquad  z_* = 0.0318 \nonumber \\
 \alpha & = & 1.769 , \qquad \beta = 4.531 , \qquad \gamma = 1.335 
     \nonumber \\
 n_* & = & 8.76 \ten{-3} \,h^{3} {\rm Mpc}^{-3} ; \nonumber 
\end{eqnarray}
these values are appropriate for $f \ge 0.6$ Jy; for other flux limits 
we simply scale $z_*$ by $\sqrt(0.6 / f_{\rm lim})$.

We have made three refinements to the FKP estimator: 
firstly, we define the ratio of
densities of real and random catalogues 
$\alpha' = \sum_g w_i / \sum_s w_j$, 
where $w_i$ is the weight of the $i$th object and 
the sums run over galaxies and random points respectively \cite{te-stromlo}, 
instead of $\alpha = N_g / N_r$ as in FKP (where
$N_g, N_r$ are the numbers of galaxies and randoms respectively). 
Secondly, we compute the shot noise using 
\begin{equation}
\label{eq-shot}
 \Pshot = \sum_g w_i^2 +  \alpha'^2 \sum_s w_j^2 , 
\end{equation}
where the two terms are the contributions from 
galaxies and random points respectively.
The shot noise definition 
in FKP Eq.~2.4.5 was $\Pshot = \alpha (1+\alpha) \sum_s w_j^2$; there
the first-order term in $\alpha$ is the 
`expected' shot noise from the galaxies given many realisations
of the given selection function, while the first term 
in our definition is the `actual' shot noise in the data. 
This makes negligible difference at large scales, 
but we find from simulations
that Eq.~\ref{eq-shot} gives substantially 
smaller errors in the estimated power spectrum at small scales (large $k$),
because the shot noise term is substantial here and the 
`actual' shot noise from the galaxies may differ significantly
from its expectation value estimated from the selection function. 

The third refinement is that we use a 
normalisation convention given by Eq.~\ref{eq-a3}; 
see Appendix~A for a discussion of the normalisation. 

The estimated power spectrum $\Phat(k)$ is then given as in FKP, by
\begin{eqnarray}
\label{eq-pest} 
 F(\br) & = & w(r) [n_g(\br) - \alpha' n_s(\br)] \\
 F(\bk) & = & \int d^3r \, F(\br) e^{i \bk.\br} \\
 \Phat(\bk) & = & \vert F(\bk) \vert^2 - \Pshot 
\end{eqnarray} 
where $n_g, n_s$ are the number densities of galaxies and random points
in cubical cells, 
and $\Phat(k)$ is just the unweighted 
average of $\Phat(\bk)$ over a spherical shell with mean radius $k$. 

The optimal weighting scheme depends on the actual value of 
$P(k)$, so the procedure is slightly circular in principle.  
We have used values of 
$\Pe = 2000, 4000, 8000, 16000 \hmpcc $; 
estimates of the power spectrum for each value of 
$\Pe$ are shown in Figure~\ref{fig-pkwt}. 

\begin{figure}
\plotfiddle{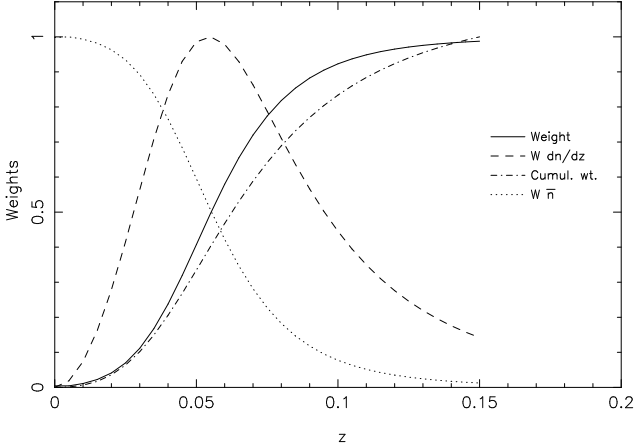}{3.0truein}{-90}{35}{35}{-10}{200} 
\caption{ Weight and window functions vs. redshift 
 for $\Pe = 8000 \hmpcc$, in arbitrary 
 units.  The solid line shows the weight $w(z)$, 
 and the dotted line shows the real-space
 window function $\nbar(z) w(z)$.
 The dashed line shows $\nbar(z) w(z) dV/dz$, i.e. the `effective
 volume' of the survey per unit redshift, and the 
 dot-dash line is the integrated effective volume below 
 redshift $z$. 
 \label{fig-weight} } 
\end{figure}

We see that changing $\Pe$
changes the size of the error bars, but there is little
systematic difference in the resulting estimates of $\Phat(k)$.
This is as expected since FKP showed that 
any choice of $\Pe$ gives an unbiased 
estimate of $P(k)$ (apart from the convolution effects at small $k$ 
discussed below), 
but just weights different redshift shells differently - larger $\Pe$
gives relatively more weight to more distant shells. 
We adopt $\Pe = 8000 \hmpcc$ as the default value for the remainder
of the paper. The resulting weights are illustrated 
in Figure~\ref{fig-weight}: the solid line shows the weight
function $w(z)$, and 
the dotted line shows the real-space window function $\nbar(z) w(z)$. 
Also shown are the differential and 
cumulative contributions to the survey `effective volume' 
per unit redshift. 

We have explored varying many of the selection criteria, 
 e.g. varying the model of the selection function, 
using a maximum redshift of $30000 \kms$, or 
a galactic latitude cut of $20^o$, and changing the flux threshold.
Most of these changes have a negligible effect on the results, 
with the exception of changing the flux limit: the subsample with 
$f_{60} > 0.8$ Jy has a slightly lower amplitude of $\Phat(k)$
at all scales; the difference is mostly within the 
individual $1\sigma$ error bars but is evident over a range of $k$. 
The slice with $0.6 < f_{60} < 0.8$ Jy has correspondingly
higher amplitude.  
This might be suggestive of systematic errors in the catalogue
near the flux limit, as suggested by Hamilton (1996); 
however we have investigated 
the correlation function $\xi(\sigma,\pi)$ as a function of projected and 
redshift separation 
and find negligible evidence for a ``Hamilton effect'' i.e. elongation
of the correlation function out to large redshift separation.  
The correlation functions of the subsamples above and below
$0.8$ Jy show a similar difference in 
amplitudes to the $\Phat(k)$'s,  but the difference is
most pronounced at small to intermediate scales, rather 
than at large scales as might be expected from systematic errors; 
thus this effect may be a sampling fluctuation or possibly 
a dependence of clustering amplitude on intrinsic galaxy luminosity. 
Hereafter we use the full sample to 0.6 Jy 
with the caveat that the reason 
for this slight flux-dependence is not yet clearly understood. 
The effect on the derived cosmological parameters in \S~\ref{sec-fits}
is similar to or less than the uncertainties arising from
the small-scale velocity dispersion etc. 

\subsection{Observed vs True P(k)} 
\label{sec-obsvtrue}

\begin{figure*} 
\plotfiddle{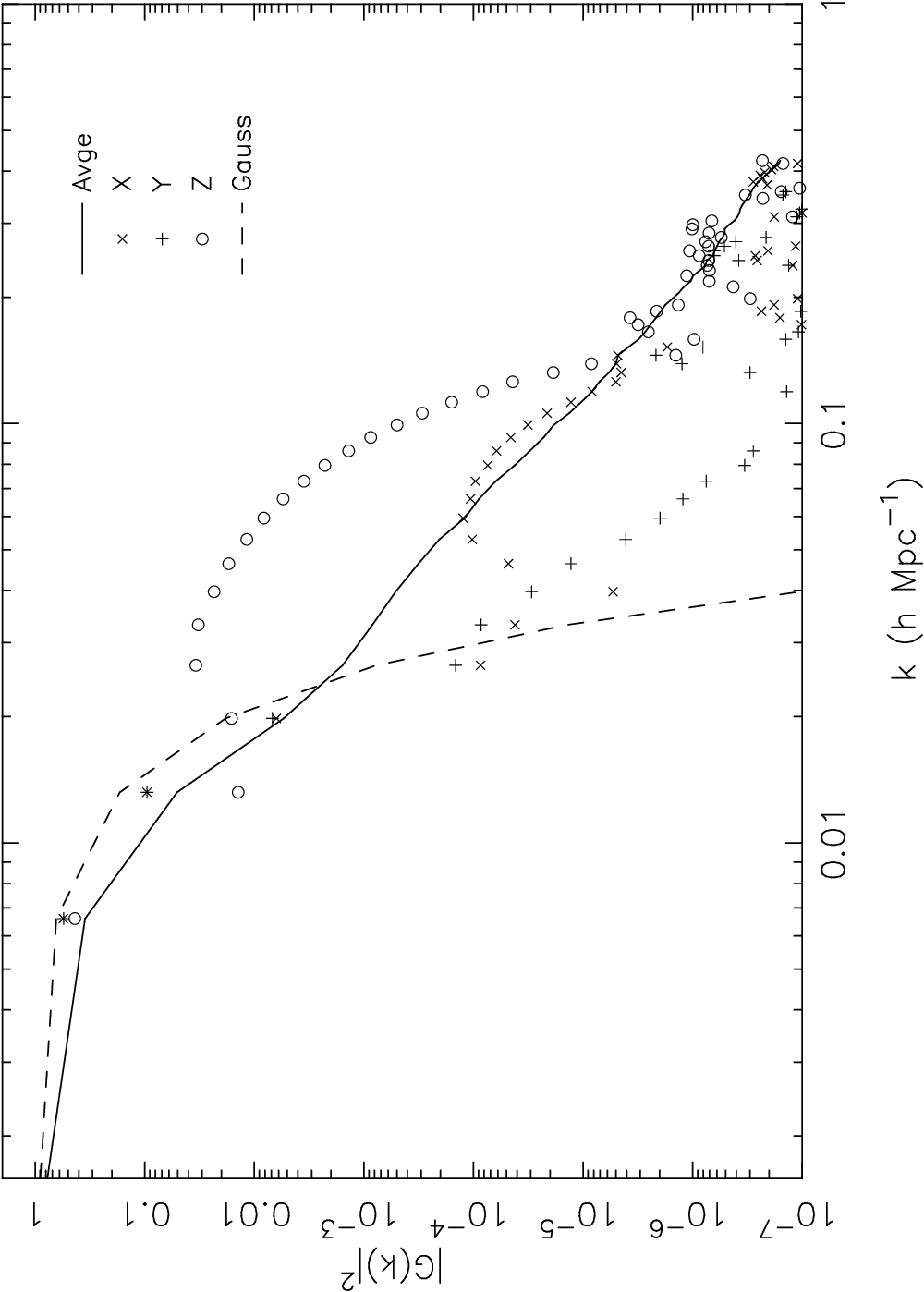}{3.1truein}{-90}{45}{45}{+80}{260} 
\caption{ The k-space window function $\vert G(\bk) \vert^2$. 
 Points  show this quantity 
 for $\bk$ parallel to the Galactic $x,y,z$ directions (as labelled).
 The solid line shows the direction-average over spherical shells
 of radius $k$. 
  The dashed line shows a Gaussian 
 $\exp(-k^2 / 2 k_0^2)$ with $k_0 = 0.007 \kunit$ (this is 
 for illustration and is not a fit). 
\label{fig-win} }
\plotfiddle{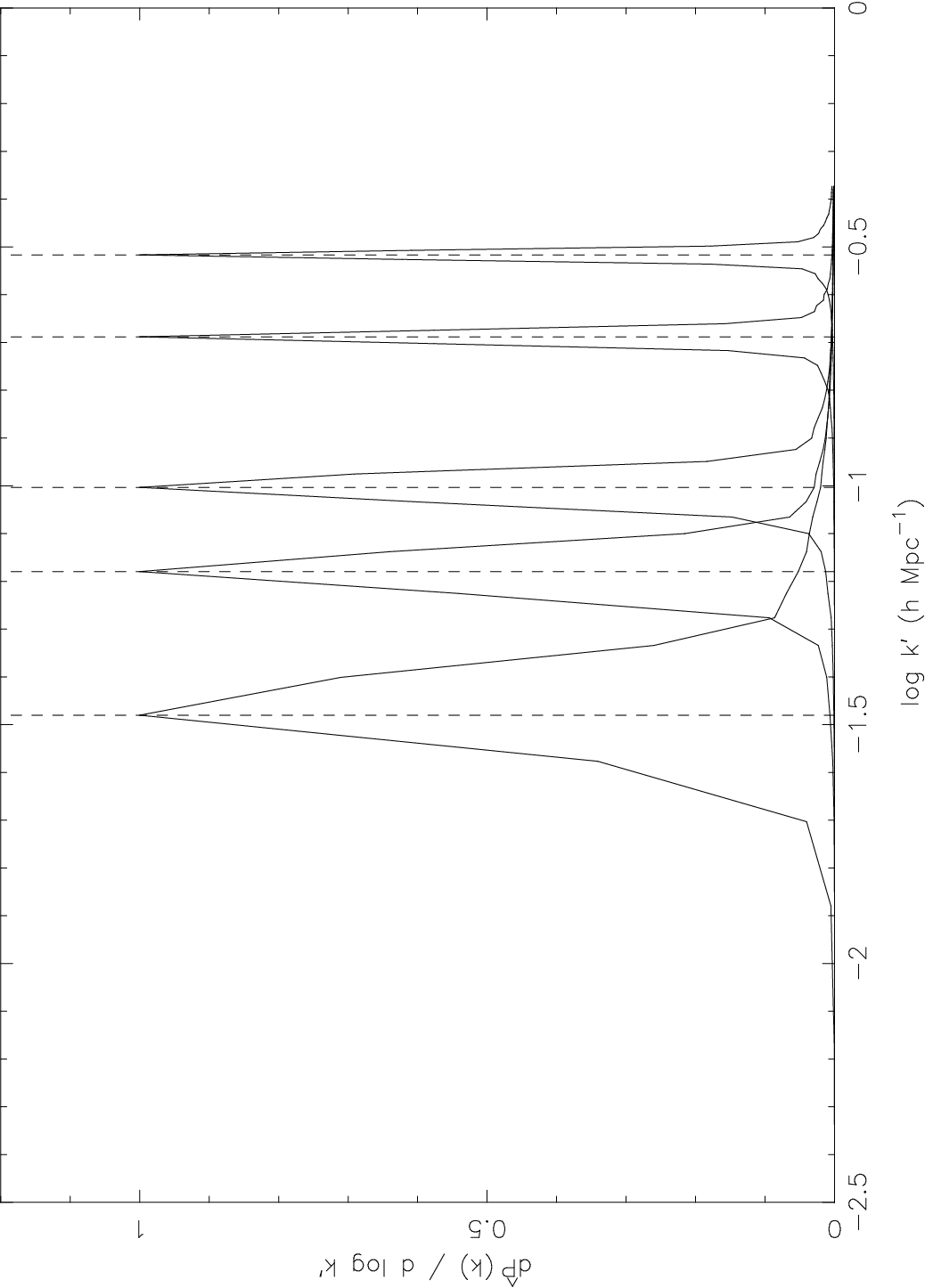}{3.1truein}{-90}{45}{45}{+80}{260} 
\caption{ The effect of convolution, i.e. the contribution 
 to `observed' power $\Phat(k)$ 
 from `true' wavenumber $k'$ in Eq.~\protect\ref{eq-convol}, 
 summed over directions of $\bk, \bk'$,  
 per unit $\ln k'$. Values are shown for observed 
 $k = 0.033, 0.066, 0.1, 0.2, 0.3 \kunit$ (dashed lines). 
 The y-scale is arbitrary.  \label{fig-conv} } 
\end{figure*}

As is well known, the finite size of the survey volume means
that the estimated power spectrum is a convolution of the true
power spectrum with the squared Fourier transform of the
real-space window function; e.g. Eqs.~2.1.6 and 2.1.10 of FKP give 
\begin{eqnarray}
\label{eq-convol}
\langle \Phat(\bk) \rangle & = & (2\pi)^{-3} \int d^3k'\, P(\bk') 
\vert G(\bk - \bk')\vert^2 \\
G(\br) & \equiv & \nbar(\br) w(\br)  \left(\int d^3r \,\nbar^2(\br) w^2(\br)
\right)^{-1/2} \nonumber , 
\end{eqnarray} 
and the normalisation is defined so that 
\begin{equation}
\int d^3k \, \vert G(\bk) \vert^2 = (2 \pi)^3
\end{equation}
by Parseval's theorem. 
This convolution is a significant problem for slice-like survey geometries
with a highly anisotropic window function; but 
since our survey is large in all 3 dimensions, the window function is narrow. 
For our standard weighting with $P_e = 8000 \hmpcc$,
the window function $\vert G(\bk) \vert^2$ 
is illustrated in Figure~\ref{fig-win}
for three axes in Galactic coordinates, and the angle-average. 
The window function
is roughly approximated by a Gaussian 
$\exp(-k^2 / 2 k_0^2)$ with $k_0 \sim 0.006 \kunit$ at small $k$,
with a roughly $k^{-4}$ tail arising from the survey mask. 
Therefore, the effect of the convolution on our estimates is small
except at the largest scales $k \simlt 0.03 \kunit$. 
This is illustrated in Figure~\ref{fig-conv}; the solid lines
show the fractional contribution to the measured $\Phat(k)$ as a function of
`true' wavenumber $k'$, i.e. Eq.~\ref{eq-convol} averaged over
directions of $\bk, \bk'$
for five values of observed $k = 0.033, 0.066, 0.1, 0.2, 0.3 \kunit$,  
assuming a CDM-like model $P(k')$ with 
the parameter $\Gamma = 0.3$ (see Eq.~\ref{eq-ebw} below). 
We have not attempted a deconvolution here, since the 
convolution effect is only important at small $k$ where the 
estimates are becoming noisy. 

\begin{figure} 
\plotfiddle{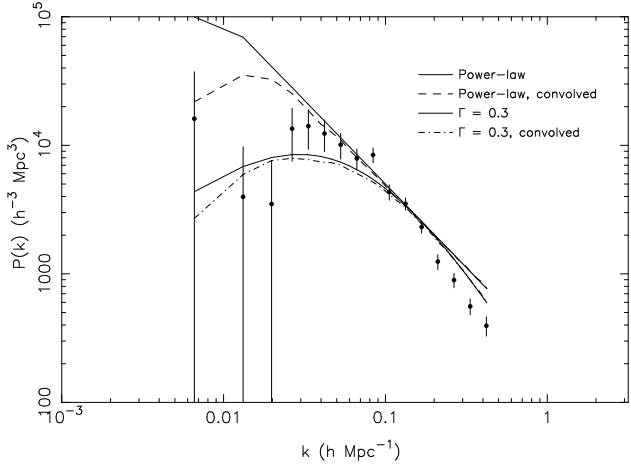}{3.0truein}{-90}{35}{35}{-10}{200} 
\caption{ The effect of convolution and the normalisation condition 
 $\Phat(k) = 0$  on two model power spectra (solid lines);  
 the upper solid line shows a power law $P(k) \propto k^{-1.2}$ 
 with a break at $k = 0.01 \kunit$; 
 the lower solid line shows $\Gamma = 0.3$ CDM. 
 The dashed and dot-dash lines show the mean recovered power spectra 
 for the two cases. Points show the measured values for PSCz. 
 \label{fig-modcon} } 
\end{figure}

Another effect which causes the measured $\Phat(k)$ to deviate
systematically from its true value is that the mean density of galaxies is not
known independently but is estimated from the survey. 
This leads to the constraint $\Phat(0) = 0$, and the convolution above means
that $\Phat(k)$ will also be underestimated for small but non-zero $k$. 
This effect was noted by  Peacock \& Nicholson (1991), 
and has been evaluated analytically by Tadros \& Efstathiou (1996)
for the special case of a  
volume-limited survey; their Eq.~A4.2 gives 
\begin{eqnarray}
\label{eq-bias}
\langle \Phat(\bk) - P(\bk) \rangle & \approx & 
  - {\nbar \over V} \vert \widehat{W}(\bk)^2 \vert \\
 & &  - \nbar^2 \left[ \sum_{\bk'}  
 \vert \widehat{W}(\bk') \vert^2 P(\bk') \right]
  \vert \widehat{W}(\bk)^2 \vert \nonumber 
\end{eqnarray} 
For unequal weights as here, the expression is complex and
best evaluated numerically;  
we have computed this for two models of $P(k)$; a $\Gamma = 0.3$ 
CDM model, and an $n = -1.2$ power-law, with a bend to $n=0$ at
 $k \le 0.01 \kunit$. 
For each model, we generate a Gaussian random density field with the
assumed $P(k)$, multiply by the PSCz window function
$\nbar(r) w(r)$,  set the mean of the windowed density field to zero, 
and compute the power spectrum of the resulting density field 
using the given $w(r)$.  
The mean of the recovered power spectra from 30 realisations of each model
is shown in Figure~\ref{fig-modcon}. 
We see that the bias is 
serious only for $k \simlt 0.02 \kunit$; note in particular that
for the power-law model, the recovered power spectrum lies well above
the PSCz measurements at $k \sim 0.01 - 0.03 \kunit$. 
The data points at $k \sim 0.013 {\rm and} 0.02 \kunit$ are noteworthy
here; although their error bars appear large due to the log scale, none of the 
30 realisations of the power-law model gave $\Phat(k)$ 
as low as the PSCz data at these $k$. 

Thus, we can be confident that the
flattening of the observed power spectrum 
below $k \sim 0.06 \kunit$
is a real feature, not an artefact of the finite volume or 
normalisation. 

We compute the covariance matrix of $\Phat(k)$ using Eq~2.5.2
of FKP. In practice, it is infeasible to evaluate this directly
since it contains $\sim N^6$ terms where $N=128$; 
thus in practice, we assign each of the $N^3$ wavevectors to its
bin in $k$; for each $k, k'$ we pick $\sim 10^6$ random pairs of
wavevectors $\bk, \bk'$ in the appropriate bins, and evaluate
the sum accordingly. It is found that $\sim 10^6$ pairs are necessary
to ensure that the random errors are small, otherwise the
resulting matrix can become non-positive-definite. 

A final effect on $\Phat(k)$ is the  `binning factor'
noted by Baugh \& Efstathiou (1994b), which 
causes an underestimate of small-scale power due to the
galaxies being binned into finite-size cells
before the Fourier transform. This `smooths' the observed
density field over the bin size. 
The size of this effect depends on the slope of the true
power spectrum; a second-order approximation is given by
Peacock \& Dodds (1996) Eq.~20, and is
$ \Phat(k) / P(k) \approx \left[ 1 + (k l)^2 / 12 \right]^{-1}$, 
where $l$ is the size of the unit cells in the FFT; this becomes
inaccurate for $ k l \simgt 2$. 
We find that for $n \sim -1$ 
a better approximation useful for $kl < \pi$ is 
\begin{equation}
\label{eq-bincor} 
\Phat(k) / P(k) \approx 
   \left[ 1 + (k l)^2 / 12 - (k l)^4 / 220 \right]^{-1}  . 
\end{equation} 

Since this correction is slightly model-dependent, 
we have not applied it to the estimates in 
Figure~\ref{fig-pkwt} or Figure~\ref{fig-psim}, but we have
applied it before the model fitting in the following Section.
 
\begin{table*} 
\caption{ Estimated power spectrum \label{tab-pk} }
\begin{center}
\begin{tabular}{@{}ccccccc} 
\hline  
\colhead{$k$} & \colhead{$\Phat(k)$} & 
 \colhead{Error} & \phantom{spaces...} & 
  \colhead{$k$} &  \colhead{$\Phat(k)$} & Error \\ 
\colhead{$(\kunit)$} & \colhead{$(\hmpcc)$} & & & 
 \colhead{$(\kunit)$}  & \colhead{$(\hmpcc)$} &  \\
\hline
0.0066 & 16110  & 21200 & &  0.0839 & 8690 & 1165 \\
0.0132 &  3981  &  5774 & &  0.1056 & 4548 &  625 \\
0.0198 &  3503  &  4039 & &  0.1329 & 3786 &  425 \\
0.0265 & 13514  &  5960 & &  0.1674 & 2591 &  285 \\
0.0334 & 14189  &  4861 & &  0.2107 & 1466 &  193 \\
0.0420 & 12459  &  3428 & &  0.2653 & 1122 &  151 \\
0.0529 & 10241  &  2321 & &  0.3339 &  748 &  119 \\
0.0666 &  8059  &  1488 & &  0.4204 &  545 &   89 \\
\hline
\end{tabular}
\end{center}
\medskip
{ This table shows the estimated power spectrum
 $\Phat(k)$ with weight $\Pe = 8000 \hmpcc$, 
 and associated $1\sigma$ errors.  
 These data points have been `corrected' for finite-size bins 
 using  Eq.~\ref{eq-bincor} with $l = 950/128 \hmpc$. 
 Covariances between consecutive entries are substantial
 for small $k$ but negligible for $k \simgt 0.1 \kunit$. }
\end{table*} 
\medskip

\section{Comparison with Simulations}
\label{sec-sim}

As a check of the code, and to assess whether the resulting error
bars are realistic, we have generated simulated `PSCz' surveys 
from large N-body simulations from 3 cosmological models
and computed their power spectra as above. 
The 3 models are `standard' CDM (SCDM), CDM with a
cosmological constant ($\Lambda$CDM), and 
a mixed cold + hot dark matter model (MDM); 
the simulations use a P3M code \cite{ce94} and 
their parameters are listed in Table~\protect\ref{tab-sims}. 
We place an `observer' at a random location in the cube, wrap the
simulation using periodic boundary conditions, and then select
`galaxies' in redshift space as random particles
 using the selection function of 
Eq.~\ref{eq-selfn} and a `mask' of the same shape as the real PSCz 
mask. 
We generated a total of 27 simulated `PSCz' surveys for each model; 
for each model we used 9 different runs of the P3M code, and
3 different observer locations for each run. 

\begin{figure} 
\plotfiddle{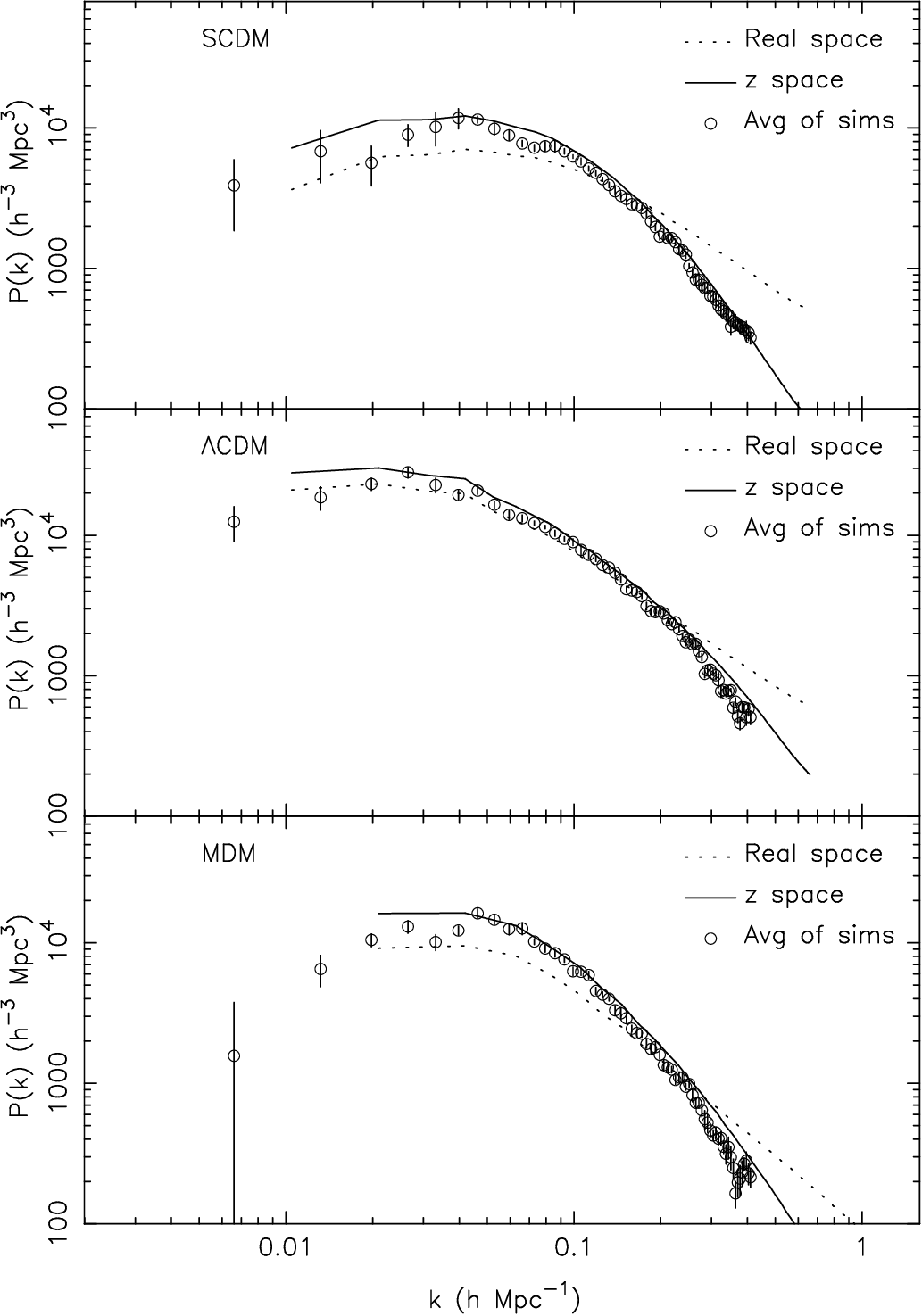}{5.0truein}{0}{45}{45}{-10}{0} 
\caption{ Power spectra for the 3 models in Table~\protect\ref{tab-sims}.
  For each model, the dotted and solid lines show the real and 
 redshift space power spectrum of the full simulation box (mean of 9 runs,
 error bars negligible).
 The circles show the mean of the estimated power spectra
 from 9 simulated `PSCz' surveys, with $1\sigma$ error on the mean.  
  \label{fig-psim} } 
\end{figure} 


\begin{table*} 
\caption{ N-body Simulation Parameters  \label{tab-sims} }
\begin{center}
\begin{tabular}{cccccccc}  
\hline\hline
\colhead{Name} & \colhead{N} & \colhead{Box ($\hmpc$)} & 
  \colhead{$\Omega_{\rm CDM}$} &  \colhead{$\Omega_{\rm HDM}$} & 
 \colhead{$\omegalam$}  & \colhead{$h$} & \colhead{$\sigma_8$} \\
\hline
SCDM         & $160^3$ & 600 & 1   & 0   & 0   & 0.5 & 1.0 \\
$\Lambda$CDM & $160^3$ & 600 & 0.2 & 0   & 0.8 & 1.0 & 1.0 \\
MDM          & $100^3$ & 300 & 0.7 & 0.3 & 0   & 0.5 & 0.67 \\
\hline
\end{tabular}
\end{center}
\end{table*} 

\begin{figure*} 
\plotfiddle{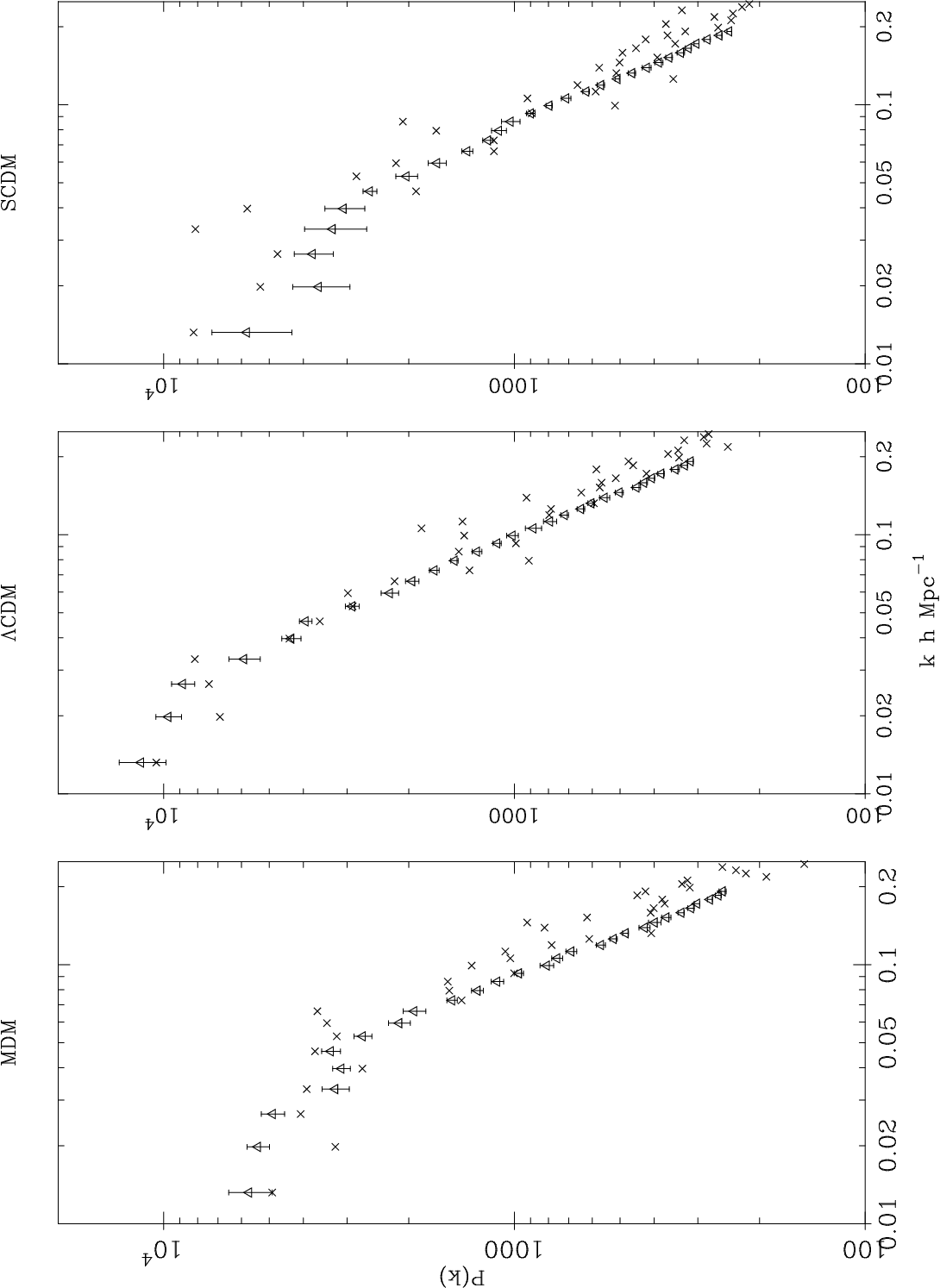}{3.5truein}{-90}{45}{45}{+90}{260} 
\caption{ Error estimates for simulated PSCz surveys, with the 
 observed mask and selection function, for the three models 
 in Table~\protect\ref{tab-sims}. 
 Triangles show the mean of the FKP error estimate $\delta\Phat(k)$ from 9 
 simulated surveys; crosses show the `true' error in $\Phat(k)$
 estimated from the scatter in the 9 simulated surveys. 
  \label{fig-errs} } 
\plotfiddle{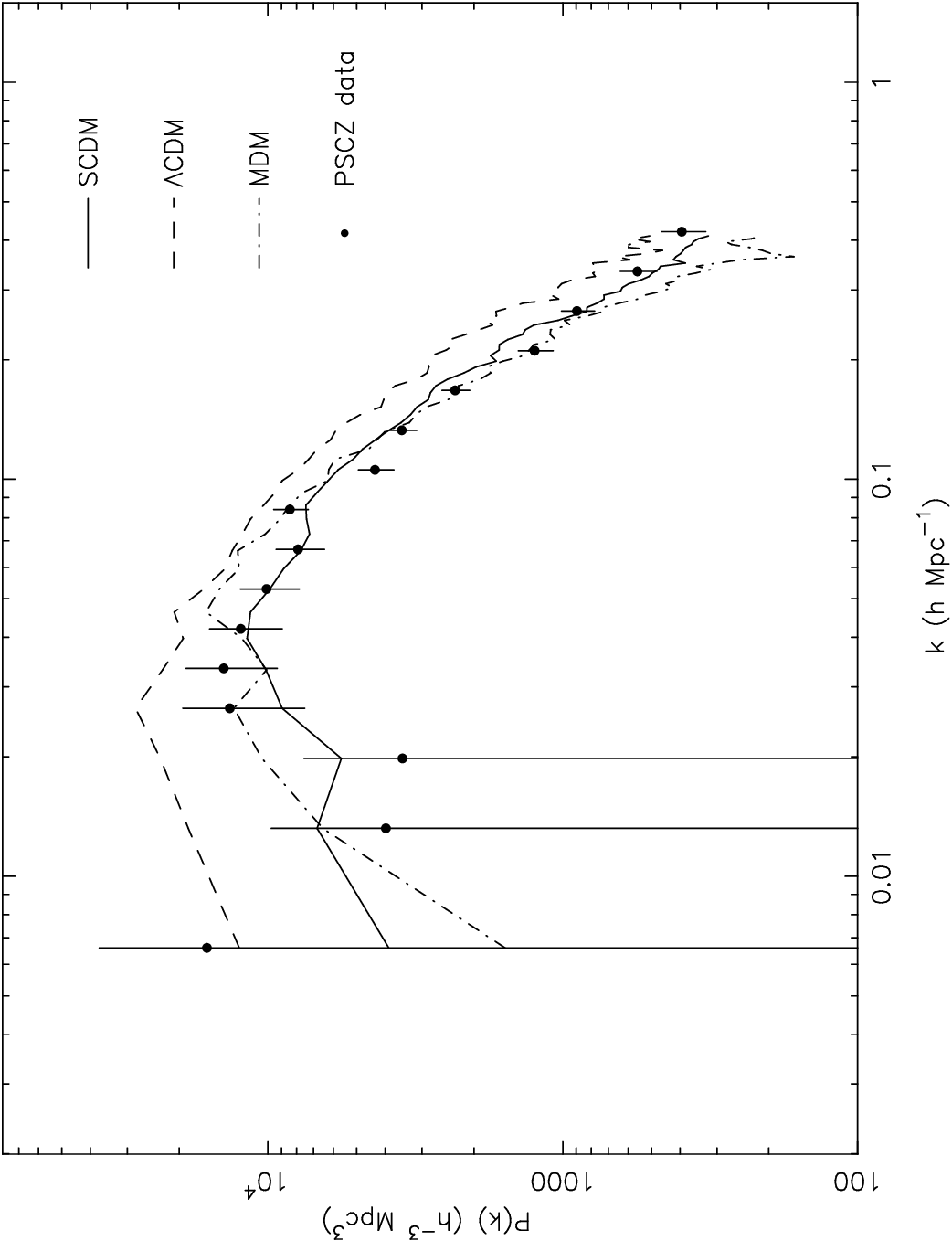}{3.5truein}{-90}{45}{45}{+90}{260} 
\caption{ The observed PSCz power spectrum (points with $1\sigma$ errors) 
  compared to the mean power spectrum  of 27 simulated PSCz surveys 
 for each of the 3 models (lines). 
  \label{fig-pdatvsim} } 
\end{figure*}

Figure~\ref{fig-psim} shows the power spectra of
the full box for each simulation, compared with results from 9
simulated PSCz surveys.  
As expected, our estimator $\Phat(k)$ recovers the `true' redshift-space
power spectrum quite well for $k \simgt 0.02 \kunit$; 
the simulated surveys slightly underestimate 
the true redshift-space $P(k)$ on intermediate scales due to
the convolution with the survey window function, but the effect is small. 
Figure~\ref{fig-errs} shows the mean of the FKP error bars from 
9 simulations (triangles), compared with the `real' uncertainty
estimated from the rms scatter between the 9 simulations (crosses). 
Clearly the FKP error estimates are a reasonable approximation
to the `real' errors in $\Phat(k)$, though they appear to 
underestimate the actual errors by $\sim 20\%$. 

In Figure~\ref{fig-pdatvsim} we compare the data to the 
mean of 27 simulated PSCz surveys for each model. 
We find that all three models give a reasonable match to
the shape of the observed power spectrum; 
the  $\Lambda$CDM model has somewhat too high an amplitude and 
would require an antibias $b < 1$. However, this could be
remedied by lowering $h$ somewhat and raising $\Omega_0$, keeping 
$\Gamma$ constant; this 
would reduce the implied $\sigma_8$ for COBE normalisation, and 
improve the fit. Alternatively, a modest degree of `tilt' with 
primordial spectral index $n < 1$ would similarly reduce the 
COBE-normalised $\sigma_8$. 

Although the SCDM model has much less 
large-scale power than the others in real space, 
the high COBE normalisation gives two effects: 
an enhancement of power on large scales 
by a factor $\approx 1.86$ from the Kaiser (1987) redshift-space distortion,
and a suppression of small-scale power from the resulting large 
peculiar velocities. These two effects combine to bring 
the redshift-space $P(k)$ of this model 
into rather good agreement with the data.
These effects for COBE-normalised SCDM have
been previously noted by various authors, e.g. Bahcall \etal\ (1993),
although of course this model has serious problems with 
cluster abundances \cite{eke-clus}, 
large-separation gravitational lenses \cite{cen-lens} etc.

\section{Discussion}
\label{sec-discuss}

\subsection{Comparison with Other Surveys} 

\begin{figure*} 
\plotfiddle{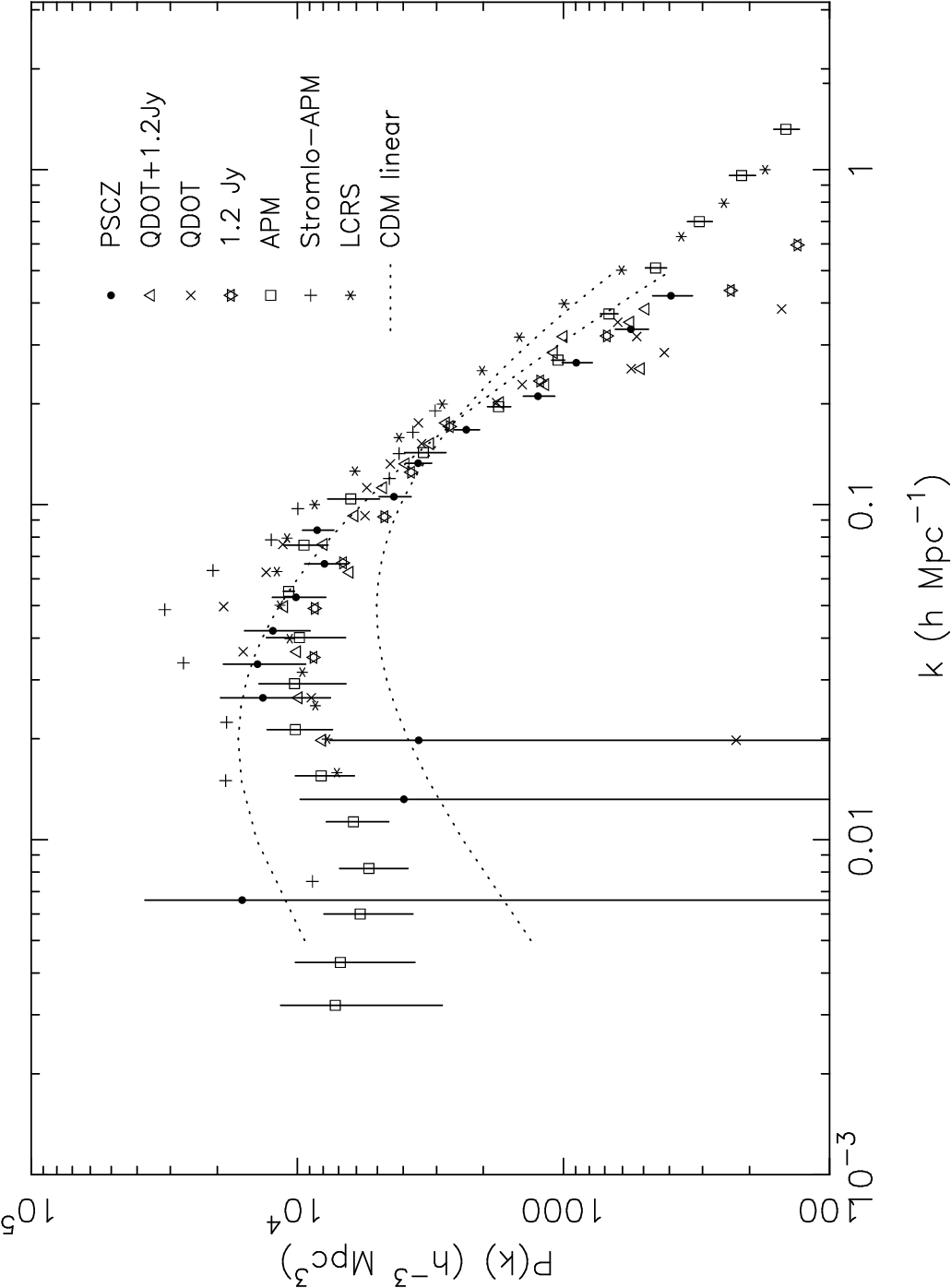}{5.0truein}{-90}{60}{60}{40}{350} 
\caption{ Comparison of PSCz $\Phat(k)$ with other 
 measured power spectra. 
 The data points as labelled on the figure are: 
 IRAS 1.2 Jansky from Fisher \etal\ (1993), 
 QDOT from FKP, 
 QDOT + 1.2 Jansky from Tadros \& Efstathiou (1995), 
 Stromlo-APM from Tadros \& Efstathiou (1996), 
 APM (real-space, deconvolved) from Baugh \& Efstathiou (1994), 
 and Las Campanas (deconvolved) from Lin \etal (1996). 
 For clarity, error bars are only shown on PSCz and APM (others are
 larger), and Stromlo, QDOT and QDOT+1.2 Jansky data have been rebinned. 
 Dotted lines show linear-theory CDM with $\Gamma = 0.2$ and 0.5 
 with $\sigma_8 = 0.8$. 
  \label{fig-surveys} } 
\end{figure*}

Our observed redshift-space power spectrum for $\Pe = 8000 \hmpcc$
is compared with a number of previous measurements in
Figure~\ref{fig-surveys}. These come from various catalogues,
both optical and IRAS selected; all are redshift-space power 
spectra except for the APM data which come from an inversion 
of the 2-D power spectrum. 
We see that on large scales our measurements are well within the
range of previous surveys, but on intermediate scales $\sim 0.2 \kunit$ 
our measurements are slightly steeper than the others, notably 
the combined QDOT and 1.2~Jy surveys. This is the cause of the low
values of $\Gamma \sim 0.2$ for best-fitting CDM-like models seen
in the next section.  
It appears that optical galaxies have a somewhat higher 
power spectrum amplitude on intermediate  scales, as expected from
the fact that IR-selected surveys contain 
mainly late-type galaxies, 
and from their smaller correlation length $r_0$,  
but it remains unclear whether this persists to large scales; 
a direct comparison with APM is complicated by the redshift-space
distortion, and the interpretation of 
LCRS is somewhat complicated by the inversion from the 2-D to 3-D power
spectrum due to the slice-like geometry.  
Future large optical surveys such as 2dF and Sloan should greatly
clarify this question. 
 
\subsection{Fits to the power spectrum} 
\label{sec-fits}

In addition to the direct comparison with simulations, it is interesting
to extract best-fitting values for parametrised models of the power
spectrum; we use firstly linear theory for simplicity, and 
later the fitting formulae of Peacock \& Dodds (1996)
which account both for non-linear evolution of clustering, and the 
effect of distortions between real space and redshift space. 

\begin{figure*}
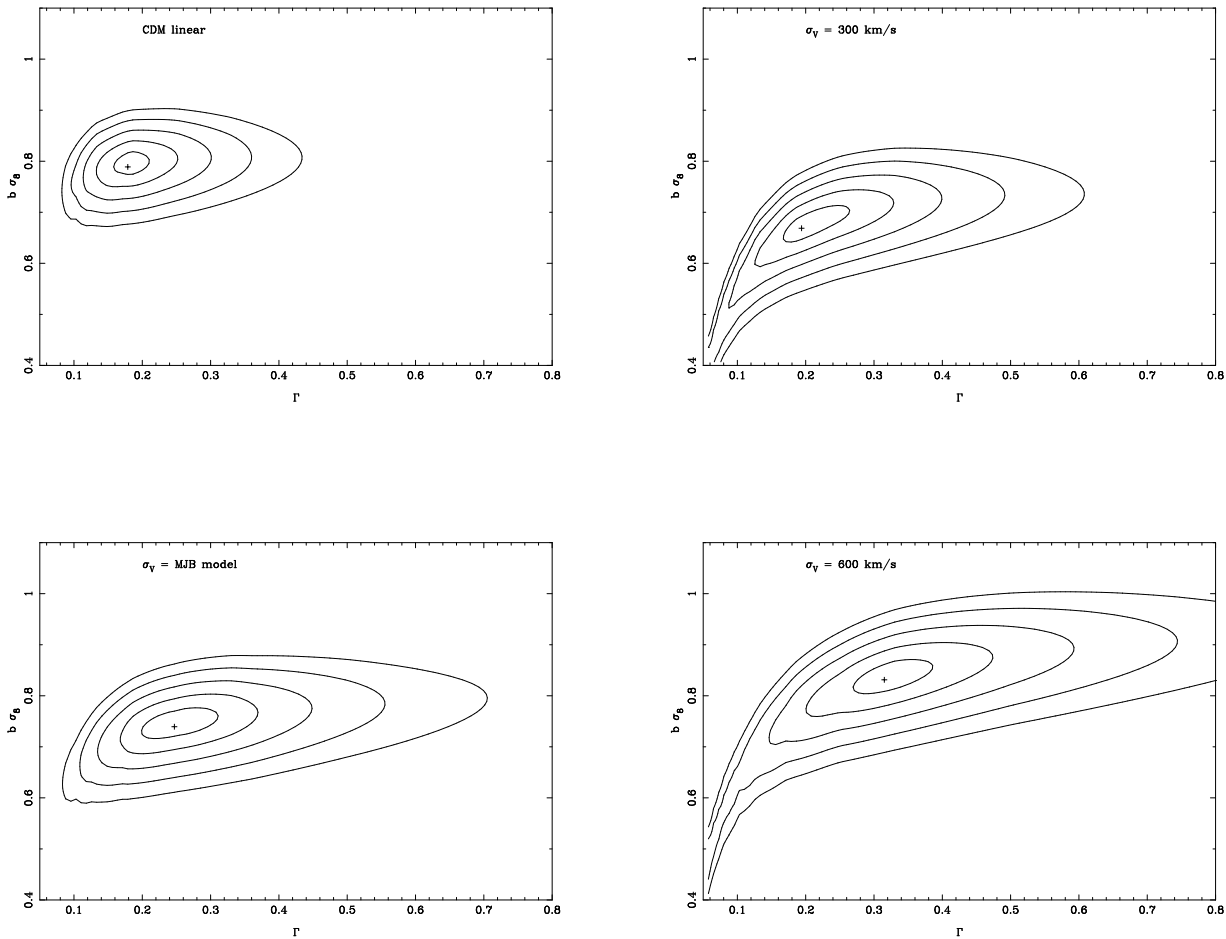
 
\plotfiddle{fig_cont_lin.ps}{2.7truein}{-90}{30}{30}{0}{200} 
\plotfiddle{fig_cont_300.ps}{-0.1truein}{-90}{30}{30}{250}{200} 
\plotfiddle{fig_cont_mjb.ps}{2.7truein}{-90}{30}{30}{0}{200} 
\plotfiddle{fig_cont_600.ps}{-0.1truein}{-90}{30}{30}{250}{200} 
\caption{ 
  Contours of $\chi^2$ in the $(\Gamma, b\sigma_8)$ plane.
  of fits to the observed power spectrum.
 The upper left panel shows linear theory. Other panels 
  use the Peacock-Dodds 
 formula as in Sec.~\ref{sec-fits}, with $\Omega = \Gamma/0.66$,
 cluster normalisation, and different
 choices of random velocities $\sigma_V$ as labelled. 
 The cross denotes the minimum $\chi^2$,
 contours are at $\chi^2 = \chi^2_{\rm min} + 1,4,9,16,25$. 
 \label{fig-cont} } 
\end{figure*}

We use CDM-like models with the initial power spectrum 
parametrised by $\Gamma$ as in 
Eq.~7 of Efstathiou, Bond \& White (1993, hereafter EBW), 
which is 
\begin{eqnarray}
\label{eq-ebw}
P(k) & = & { B k \over 
   \left( 1 + [ak + (bk)^{3/2} + (ck)^2 ]^\nu \right)^{2/\nu} } \\
  a & = & 6.4 / \Gamma \hmpc, \quad b = 3.0/\Gamma \hmpc, \nonumber \\
  c & = & 1.7 / \Gamma \hmpc \quad \nu = 1.13 \nonumber
\end{eqnarray}
 In linear theory there are only 2 free parameters: the
shape parameter $\Gamma$, and the normalisation which may be taken
as $b \sigma_8$, where $b$ is the bias parameter and 
$\sigma_8$ is the rms mass fluctuation in an $8 \hmpc$ top-hat sphere.
The $\chi^2$ contours using the full covariance matrix are
shown in Figure~\ref{fig-cont}; 
the best fit values are $\Gamma = 0.19 \pm 0.03$
and $b \sigma_8 = 0.80 \pm 0.02$. 
These compare to $\Gamma = 0.19 \pm 0.06, 
 b \sigma_8 = 0.87 \pm 0.07$ for the QDOT survey, as 
given by FKP Eq.~4.3.3.    

For the non-linear Peacock-Dodds formula, we need a total of
6 parameters to specify the present-day redshift space power spectrum: 
the initial mass power spectrum is specified as above
by $\Gamma$ and $\sigma_8$
(where $\sigma_8$ is defined as the {\sl initial} rms mass
fluctuation, multiplied
 by the linear-theory growth factor to the present day). 
The subsequent non-linear evolution depends also
on $\Omega_0$ and (weakly) on $\omegalam$.
The transformation from real to redshift space
depends on $\Omega_0$ and the bias $b$, 
and also on the pairwise peculiar
velocity dispersion; in the PD formula this is 
approximated by assuming galaxy velocities are independent 
Gaussians with dispersion $\sigma_V = \sqrt{v_{12}^2 (r) / 2}$ 
for some suitable scale $r$.  
We assume simple `linear bias' so the galaxy power 
spectrum is $b^2 \times$ the matter power spectrum.  
There is clearly insufficient information in the $P(k)$ data to
fit all 6 parameters separately, so we restrict the parameter space as
follows:

\begin{table*} 
\caption{ Fits of CDM-like models \label{tab-fits} }
\begin{center}
\begin{tabular}{cccccc}  
\hline\hline
\colhead{$\Omega$} & \colhead{Norm ($\sigma_8$) } & 
 \colhead{$\sigma_V (\kms)$} 
 & \colhead{$\Gamma$} &   \colhead{$b \sigma_8$} & \colhead{Plot?}  \\
\hline
Linear &  --- & ---          &  0.19  & 0.80 &  * \\ 
$1$           & COBE (0.45) &  300        & 0.20  & 0.67 & \\
$1$           & COBE (0.49) &  MJB (368)  & 0.22  & 0.68 &  \\
$1$           & COBE (1.00) &  600        & 0.40  & 0.71 & \\
$1$           & Clus (0.52) &  300        & 0.20  & 0.66 & \\
$1$           & Clus (0.52) &  MJB (390)  & 0.23  & 0.69 & \\
$1$           & Clus (0.52) &  600        & 0.34  & 0.81 & \\
$\Gamma/0.66$ & COBE (0.97) &  300        & 0.16  & 0.66 & \\
$\Gamma/0.66$ & COBE (0.97) & MJB (457)   & 0.16  & 0.70 & \\
$\Gamma/0.66$ & COBE (1.38) &  600        & 0.34  & 0.73 & \\
$\Gamma/0.66$ & Clus (0.91) &  300        & 0.20  & 0.67 & * \\
$\Gamma/0.66$ & Clus (0.81) & MJB (438)   & 0.25  & 0.74 & * \\
$\Gamma/0.66$ & Clus (0.73) &  600        & 0.31  & 0.83 & * \\
\hline
\end{tabular}
\end{center}

{ Fits to the observed $\Phat(k)$  in the $(\Gamma, b\sigma_8)$ plane.
 The first line shows linear theory, the remainder use
 the non-linear PD formula, 
 using various choices for setting the other
 input parameters ($\Omega, \omegalam, \sigma_8, \sigma_V$) 
 as a function of $\Gamma$. Column 1: either $\Omega = 1, \omegalam = 0$ 
 or $\Omega = 0.66/\Gamma, \omegalam = 1 - \Omega$; 
 Column 2: $\sigma_8$ defined either by COBE or cluster normalisation; 
 Column 3: $\sigma_V$  either fixed to 300 or $600 \kms$, 
 or defined as a function of the other parameters using the MJB model. 
 For the $\sigma_8$ and $\sigma_V$ columns, the values in brackets
 show the derived values at the best-fitting $\Gamma$. 
 The * in the last column flags the models plotted in 
 Figs.~\protect\ref{fig-fits} and \protect\ref{fig-cont}. }
\end{table*} 

(a) To set $\Omega$, we consider either Einstein-de Sitter models, 
with $\Omega = 1$, $\omegalam = 0$, treating $\Gamma$ as
a free parameter (which is a reasonable approximation to e.g. mixed 
 dark matter models); or we consider $\Lambda$CDM models
with $\Omega = \Gamma / 0.66$, $\omegalam = 1 - \Omega$ 
(i.e. assuming $\Gamma = \Omega h$ with a Hubble constant $h = 0.66$, 
consistent with most recent measurements);  

(b) To set $\sigma_8$, we use either 
the cluster normalisation $\sigma_8 = 0.52 \Omega^{(-0.52 + 0.13 \Omega)}$ 
\cite{eke-clus}, or we use COBE normalisation 
where $\sigma_8$ is a function of 
$\Gamma$ using Eq.~6 of EBW and $Q_{\rm rms} = 17 \mu K$. 

(c) To set $\sigma_V$, we fix either $\sigma_V = 300$ 
or $600 \kms$, or
predict $\sigma_V$ as a function of $\Gamma,\sigma_8$ 
using the fitting formula in Eqs.~40a, 40b of Mo, Jing \& Borner (1997,
hereafter MJB). A value as high as $600 \kms$ is probably disfavoured
by observations \cite{lsb}, but we include this as a conservative upper limit
to show the effect on the derived parameters. 

Having made one choice from each of (a),(b),(c) above, this defines 
$\Omega, \omegalam, \sigma_8, \sigma_V$ as a function of $\Gamma$; 
we then treat $\Gamma$ and $b$ as free parameters, and fit to
the observed $\Phat(k)$ data. 
We choose to use only the data points in the range 
 $ 0.021 < k < 0.3 \kunit $ in the fits, since points at lower $k$
may be affected by the convolution, and those at higher $k$
are subject to large and somewhat uncertain 
corrections both for the peculiar velocity term
and the binning correction of Eq.~\ref{eq-bincor}. 

\begin{figure} 
\plotfiddle{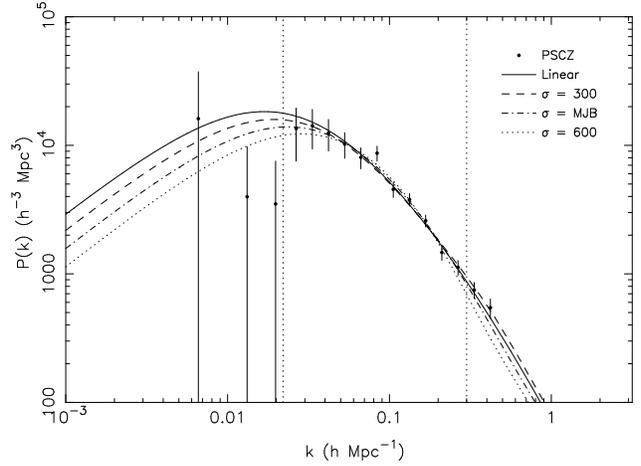}{3.0truein}{-90}{35}{35}{-10}{200} 
\caption{ 
  Fits to the measured power spectrum (points) 
 using the Peacock-Dodds 
 formula as in Sec.~\ref{sec-fits}. 
 Vertical dotted lines denote the range of $k$ used in the fits. 
 Lines show the fitted power spectra as labelled,
 one for linear theory and three for different values
 of $\sigma_V$, for the case of a low-density
 universe with $\Omega = \Gamma/0.66$, and cluster normalisation.
 \label{fig-fits} } 
\end{figure} 

The best-fitting values of $\Gamma, b\sigma_8$ for each of
the above parameter choices are shown in Table~\protect\ref{tab-fits}. 
(We present the fit results in terms
of $(\Gamma, b\sigma_8)$ rather than $(\Gamma, b)$
because $b$ and $\sigma_8$ are degenerate in the linear regime, 
and also for simplicity 
since the contours of equal $\chi^2$ are roughly parallel to the 
axes in the $(\Gamma, b\sigma_8)$ plane.)
For the low-$\Omega$ cluster-normalised 
case, contours of goodness of fit are shown in Figure~\ref{fig-cont}  
for each of the three $\sigma_V$ assumptions.
The derived power spectra for the best fits
in the same cases are shown in Figure~\ref{fig-fits}.  
For most of the fits, a fairly small value of $\Gamma \sim 0.2$ is
favoured, and $\Gamma \sim 0.5$ is quite strongly ruled out; 
for the cases with $\sigma_V = 600 \kms$, higher
values $\Gamma \sim 0.35$ are favoured
and $\Gamma = 0.5$ is only marginally ruled out,  
though values of $\sigma_V$ as high as this are definitely
not favoured observationally.  
 
 There is rather little difference 
between the goodness-of-fit for the
various choices, though the fits for $\sigma_V=600 \kms$ are somewhat worse.  
For a given parameter choice, the $1\sigma$ random errors
are typically $\pm 15\%$ in $\Gamma$ and $\pm 4\%$ in $b \sigma_8$.  
In general we see that the systematic uncertainties from the
different choices of $\Omega, \sigma_8, \sigma_V$ 
dominate the random errors arising from the error bars on $\Phat$; 
the largest source of uncertainty is that arising from $\sigma_V$. 
As expected, if we increase $\sigma_V$ the predicted 
small-scale power decreases at a fixed $\Gamma$, 
and thus the best-fitting values of $\Gamma$ and
$b\sigma_8$ increase to compensate. 

For $\Lambda$CDM models, the value of
$\Gamma$ at which the COBE and cluster
normalisations agree is $\Gamma = 0.17, \sigma_8 \approx 1.0$; 
this is interestingly close to our best-fit values 
of $\Gamma \sim 0.2$. 
Our fit values of $b \sigma_8$ thus imply
$b \sim 0.75$:  this is a significant amount of ``antibias'', but
may be plausibly accounted for by the
deficiency of IRAS galaxies in rich clusters. 
Thus a $\Lambda$CDM model is attractive in that it can simultaneously
satisfy three constraints (COBE, cluster abundance, $P(k)$) 
with one free parameter
$\Gamma$, if optical galaxies are approximately unbiased
and IRAS galaxies mildly antibiased relative to the mass. 

\subsection{Periodicities and Spikes} 

There have been a number of suggestions of a `preferred scale' for
large-scale clustering, notably by Broadhurst \etal\ (1990), 
Landy \etal\ (1996), and Einasto \etal\ (1997). 
These effects, if real, could arise
from a `spike' in the power spectrum, such as may arise from a 
baryon isocurvature model or non-standard inflation models;
or from non-Gaussian initial conditions, 
which could lead to one particular direction showing a value
of $ \vert \delta(\bk) \vert^2$ much larger than its expectation value
$P(k)$; or even from an intrinsic `preferred direction' in the Universe. 

In our data, there is marginal evidence for a `spike',  perhaps better
described as a `step', in the power
spectrum near $k \sim 0.08 \kunit$, but this is only about a $2\sigma$
effect above a smooth CDM-like fit, and the scale is significantly different
from that $k \sim 0.06 \kunit$ suggested by Broadhurst \etal\ (1990)
and Landy \etal\ (1996); also, we find that changing our flux limit to
$f_{60} > 0.7$ Jy causes a substantial drop in $\Phat(k)$ at
 this point, while leaving other points virtually unchanged. 
Thus, we suspect this point may be a statistical fluctuation, 
and there is no conclusive 
evidence for a feature in our power spectrum.
From inspection of results from a number of realisations of
N-body simulations, we find that such features quite commonly 
arise from statistical fluctuations.

\begin{figure*} 
\plotfiddle{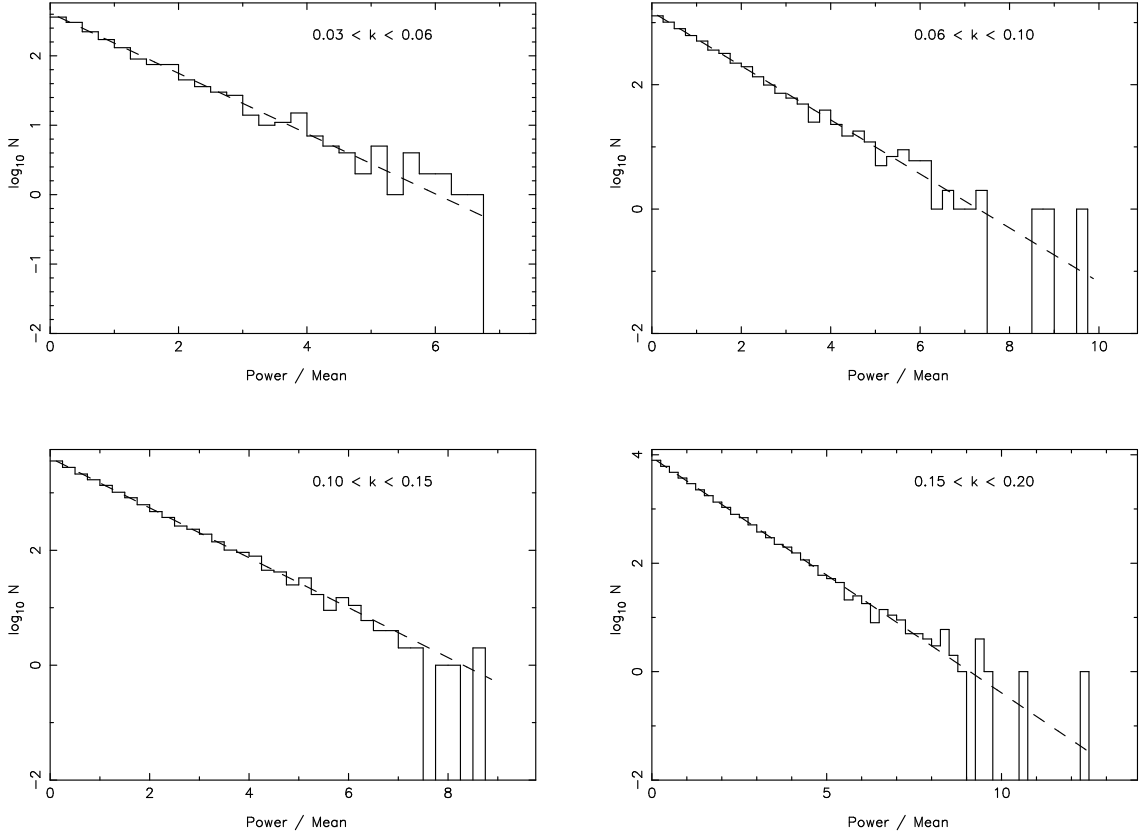}{5.0truein}{-90}{60}{60}{30}{350} 
\caption{ Solid lines show histograms of 
 `observed' power $\Phat(\bk) + \Pshot$ 
 divided by the mean for each $k$. 
 The dashed line shows an exponential distribution with unit mean. 
 The four panels show various ranges in $k$ as labelled. 
  \label{fig-pdis} } 
\end{figure*}

To test for non-Gaussian effects, we have examined the histogram
of the ratio of observed to mean power \\ 
$\vert F(\bk) \vert^2 / (P(k) + \Pshot)$ 
for each wavenumber; for Gaussian clustering, this should follow
an exponential distribution with unit mean. 
Results for several ranges of wavenumber are 
shown in Figure~\ref{fig-pdis}, and closely follow the exponential
distribution. The presence of the shot noise somewhat weakens
this test, since for smaller scales $\Pshot \simgt P(k)$, 
but at large scales this is a sensitive test for non-Gaussian 
initial conditions with a `tail' to high values. 
The distributions are well fitted by the exponential, 
so there is no evidence for non-Gaussian 
initial conditions or any `preferred direction' in our survey. 

This also strongly constrains any `preferred direction', as
follows. If there exists a strict plane-wave periodicity in the
universe with dimensionless density contrast $a$ along wavevector $\bk_0$, 
the true power spectrum contains delta functions at $\pm \bk_0$, 
$P(\bk) = C[\delta(\bk -\bk_0) + \delta(\bk + \bk_0)] $. 
Requiring the variance $\sigma^2 = a^2 / 2 
  = (2 \pi)^{-3} \int d^3k \, P(\bk)$
gives the constant $C = (2 \pi)^3 a^2 / 4$. 
Then, after convolution with the survey window function as in
Eq.~\ref{eq-convol}, 
this contributes to the observed power spectrum a term of size 
\begin{equation} 
\label{eq-spike}
\Phat_{spike}(\bk) = {a^2 \over 4} \, \left( \vert G(\bk - \bk_0) \vert^2 
  + \vert G(\bk + \bk_0) \vert^2 \right) \,. 
\end{equation}
Recall that 
\begin{equation}
\label{eq-g0} 
\vert G(k = 0) \vert^2 = { \left( \int d^3r \, \nbar(\br) w(\br)
 \right)^2 \over \int d^3r \, \nbar^2(\br) w^2(\br)  } \, ; 
\end{equation}
for a volume-limited  survey with constant $\nbar, w$ 
this just reduces to the survey volume $V$, 
so it may be thought of as the survey `effective volume'. 
For our survey with weight function 
given by $\Pe = 8000 \hmpcc$, we have 
$\vert G(0) \vert^2  = 6.2 \ten{7} \hmpcc$: thus 
even a small amplitude periodic wave of $a = 0.15$ would lead to a large 
spike in our measured $\Phat(\bk_0) \approx 3 \ten{5} \hmpcc$,  
well outside the exponential tail of measured values. 
We conclude that there is no strictly periodic plane-wave 
structure in our survey volume with amplitude larger
than $15\%$.

\section{Conclusions} 

Our conclusions may be summarised as follows: 

(i) The redshift-space power spectrum of the PSCz survey is intermediate
 between those of the earlier QDOT and 1.2 Jansky IRAS surveys on large
 scales, though it is slightly steeper on small scales. It is stable
 against variations in the galactic cuts and redshift limits etc, 
 though the amplitude decreases slightly for a flux cut $f \simgt 0.8$
 Jy.  
 
(ii) There is convincing evidence for curvature in the 
 power spectrum; the slope changes 
 from the small-scale power law $n \approx -1.4$ to 
 $n \sim 0$ on scales  $k \simlt 0.07 \kunit$. 
 This is not an artefact of the finite sample volume or the 
 estimation of the mean density from the survey. 

(iii) The best-fitting CDM-like models have $\Gamma \sim 0.25, 
 b \sigma_8 \sim 0.7$. The uncertainties in these values are
 mainly due to uncertainties in the small-scale velocity dispersion, 
 the value of $\Omega$ etc, rather than statistical errors.  

(iv) There is little evidence for a `spike' in the power spectrum, 
 and no evidence for large-scale periodicity or non-Gaussianity.

\section*{Acknowledgements}

We are very grateful to many observers, especially Marc Davis, Tony Fairall, 
Karl Fisher \& John Huchra for supplying redshifts 
in advance of publication. 
We thank the staff at the INT, AAT and CTIO telescopes for support, 
especially Hernan Tirado and Patricio Ugarte at CTIO. 
W.~Sutherland is supported by a PPARC Advanced Fellowship. 
W.~Saunders is supported by a Royal Society Fellowship. 
GE thanks PPARC for the award of a Senior Fellowship. 

\begin{appendix} 
\section{Normalisation of P(k)} 

We note an issue concerning the overall normalisation of $P(k)$. 
FKP set a normalisation of their weight function 
via Eq.~2.4.1. 
It is convenient in practice to set weights via Eq.~\ref{eq-weight} 
with $A = 1$, 
and then later divide all power spectra by a constant $A$. 
There are several ways of doing this, which give 
similar but not identical results: 
The LHS of FKP Eq.~2.4.1 gives 
\begin{equation} 
A_1 = \int d^3r \, \nbar^2(r) w^2(r) ; 
\end{equation} 
the RHS gives 
\begin{equation} 
\label{eq-a2}
A_2 = \alpha \sum_s \nbar(r_s) w^2(r_s) , 
\end{equation} 
where the sum runs over random points, and $\alpha = N_g / N_r$ 
as before. 
Another possible definition is 
\begin{equation} 
\label{eq-a3}
A_3 = {\alpha^2 \over v} \left(\sumcel c_i^2 - \sum_s w_i^2 \right), 
\end{equation} 
where the first sum runs over the cells used in the Fourier transform, 
$c_i$ is the sum of weights of all random points in the $i$th cell
and $v$ is the volume of a unit cell. 
(This results from estimating the window function from the FFT of the 
random points via 
\begin{equation}
 \vert \widehat{G}(\bk) \vert^2 = {\alpha^2 \over A_3 V} \left( \left| \sumcel 
  c_i e^{i \bk.\br_i} \right|^2 - \sum_s w_i^2 \right)  \ ; \nonumber 
\end{equation}  
requiring $\sum_\bk \vert \widehat{G}(\bk) \vert^2 = 1$ leads to
Eq.~\ref{eq-a3}.) 

The relationship between $A_1, A_2, A_3$ is as follows: 
the number of random points $N_r$ is arbitrarily fixed, and we define 
$\beta = \int d^3r \, \nbar(r) / N_r$ to be the ratio 
of the {\sl expected} number of galaxies from the selection
function to the number of randoms, which  
should be similar but not identical to $\alpha$.  
Thus the expected number
of randoms in cell $i$ is just $\nbar_i v / \beta$ where 
$\nbar_i \equiv \nbar(\br_i)$. 
For small cells, all randoms in a given 
cell will have equal weight $w_i$, 
thus $c_i = w_i \times {\rm Poisson}(\nbar_i v / \beta)$. 
Thus 
\begin{eqnarray}
\langle A_2 \rangle & = & \alpha \sumcel 
     \langle c_i \rangle \nbar_i w^2_i  \nonumber \\
     & = & \alpha \sumcel \nbar^2_i v w^2_i / \beta \nonumber \\
     & = & {\alpha \over \beta} A_1  
\end{eqnarray} 

Since $c_i$ is a Poisson variable multiplied by $w_i$, 
$\langle c_i^2 \rangle = w_i^2 [(\nbar_i v / \beta)^2 + \nbar v /\beta]$, 
thus Eq.~\ref{eq-a3} gives
\begin{eqnarray} 
\langle A_3 \rangle & = & {\alpha^2 \over v} 
  \left( {v^2 \over \beta^2} \sumcel w_i^2
    \nbar_i^2 \nonumber \right. \\
    & & \left. + {v \over \beta} \sumcel w_i^2 \nbar_i 
     - \sumcel (\nbar_i v / \beta) w_i^2 \right) \nonumber \\
   & = & {\alpha^2 v \over \beta^2} \sumcel w_i^2 \nbar_i^2 \nonumber \\
   & = & {\alpha^2 \over \beta^2} A_1  
\end{eqnarray}

Thus we see that the three definitions of $A$ differ by
powers of $\alpha/\beta$, which is just the ratio 
of `observed' to `expected' number of galaxies. 
Which is more `correct' is largely a matter of choice, 
but the definition $A_3$ is convenient 
in practice since it leaves the results unchanged 
if we  rescale the selection
function by a constant $\nbar \rightarrow c \nbar$ and 
rescale the weights by $\Pe \rightarrow \Pe / c$.  

\end{appendix}
 


\end{document}